\documentclass[conference]{IEEEtran}
\IEEEoverridecommandlockouts

\usepackage{cite}
\usepackage{amsmath,amssymb,amsfonts}
\usepackage{algorithm}
\usepackage{algpseudocode}

\usepackage{graphicx}
\usepackage{textcomp}
\usepackage{xcolor}
\usepackage{multirow}
\usepackage{subcaption}
\usepackage{amsmath}

\def\BibTeX{{\rm B\kern-.05em{\sc i\kern-.025em b}\kern-.08em
    T\kern-.1667em\lower.7ex\hbox{E}\kern-.125emX}}

\begin{document}

\title{Cell-free ISAC for Drone Detection Considering Coverage and Age of Sensing \\

\author{\IEEEauthorblockN{Zinat Behdad, Ozan Alp Topal, and Cicek Cavdar} 
\IEEEauthorblockA{Department of Computer Science, KTH Royal Institute of Technology, Stockholm, Sweden \\ {Email: \{zinatb, oatopal, cavdar\}@kth.se}}
}
\thanks{This work was supported by Swedish Innovation Agency (VINNOVA) funded through the SweWIN center (2023-00572).}
}
\maketitle

\begin{abstract}
The growing presence of unauthorized drones poses significant threats to public safety, underscoring the need for aerial surveillance solutions. This work proposes a cell-free integrated sensing and communication (ISAC) framework enabling  drone detection within the existing communication network infrastructure, while maintaining communication services. The system exploits the spatial diversity and coordination of distributed access points (APs) in a cell-free massive MIMO architecture to detect aerial passive targets. To evaluate sensing performance, we introduce two key metrics: age of sensing (AoS), capturing the freshness of sensing information, and sensing coverage. The proposed AoS metric includes not only the transmission delays as in the existing models, but also the processing for sensing and networking delay, which are critical in dynamic environments like drone detection. We introduce an ambiguity parameter quantifying the similarity between the target-to-receiver channels for two hotspots and develop a novel network configuration strategy, including hotspot grouping, AP clustering, and sensing pilot assignment, leveraging simultaneous multi-point sensing to minimize AoS. Our results show that the best trade-off between AoS and sensing coverage is achieved when the number of hotspots sharing the same time/frequency resource matches the number of sensing pilots, indicating ambiguity as the primary factor limiting the sensing performance. 
\end{abstract}

\vspace{-2mm}
\section{Introduction}
The increasing incidence of unauthorized drones poses serious threats to public safety, privacy, and critical infrastructure, highlighting the need for efficient and resilient aerial surveillance solutions. Integrated sensing and communication (ISAC) enables wireless systems to simultaneously detect and track aerial targets while maintaining seamless communication services. In this context, the distributed architecture of cell-free massive multi-input multi-output (MIMO) systems offers distinct advantages for ISAC implementation, including enhanced coverage, improved spatial diversity, and higher detection accuracy, making them particularly suitable for detecting unauthorized drones \cite{cell-free-book, zheng2021uav, wcnc2025UAV}. 

For aerial targets, maintaining high detection probability and low false alarm probability across a wide area is critical for reliable sensing. Accordingly, we define sensing coverage as the percentage of the area where the detection probability exceeds a specified threshold. Beyond accuracy and coverage, the timeliness of sensing is crucial for detecting drones before they leave the monitored area. To quantify this, we adopt the Age of Sensing (AoS), which measures the freshness of sensing data by tracking the time elapsed since the data was generated \cite{zheng2024average}. AoS inherently depends on the time required to update sensing information.

In \cite{Jing2023Peak}, beamforming, resource allocation, and offloading of the tasks are optimized to minimize peak age of information (PAoI) in Internet of Vehicles (IoV) networks. However, the work does not consider ISAC or cell-free massive MIMO systems. The feasibility of deploying ISAC systems over current and future wireless networks for aerial target detection has been studied in several works. In \cite{jopanya2025utilizing}, a sensing mechanism based on the synchronization signal block (SSB) in 5G networks is proposed, focusing on passive bi-static sensing. The authors in \cite{jopanya2025enabling} investigate the role of repeaters in enhancing radar sensing for drone detection within a swarm repeater-assisted MIMO ISAC system. However, these works do not consider the timeliness of sensing information. 
While AoS and sensing coverage for aerial target detection are studied in \cite{wcnc2025UAV} within a cell-free massive MIMO ISAC framework, the proposed AoS metric accounts only for transmission delays.

In this paper, we propose a closed-form expression for AoS that accounts not only for transmission delays but also for processing and networking delays. To reduce AoS, the surveillance area is divided into non-overlapping sensing hotspots, which are then grouped to leverage simultaneous multi-point sensing. To ensure signal separation, we employ a set of orthogonal sensing pilot sequences, allowing signals from different locations to remain orthogonal at the serving receive access points (RX-APs). Furthermore, to maximize sensing coverage, we define an ambiguity parameter that quantifies how similar the target-to-RX-AP channels are between two hotspots. Building on this, we propose a novel network configuration strategy involving AP clustering, hotspot grouping, and sensing pilot assignment. The pilot assignment is formulated as an optimization problem that aims to minimize the maximum sum ambiguity across the network.

\section{System Model}

\begin{figure}[t]
\vspace{4mm}
\centerline{\includegraphics[trim={1mm 0mm 0mm 0mm},clip,width=0.6\linewidth]{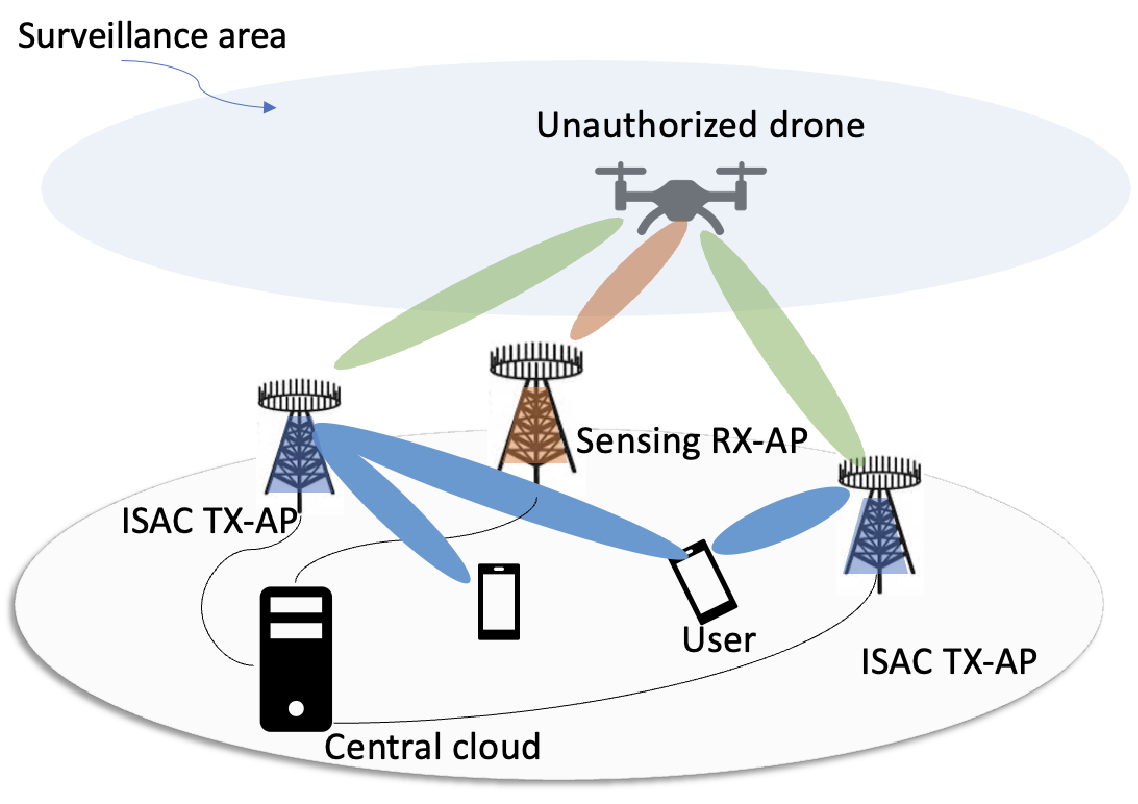}}
\caption{Aerial surveillance system with cell-free ISAC}
\label{fig1}
\vspace{-4mm}
\end{figure}

As illustrated in Fig.~\ref{fig1}, we consider downlink communication and multi-static sensing--where sensing transmitters and receivers are not co-located-- in an ISAC system within a cell-free massive MIMO setup on top of the centralized radio access network (C-RAN) architecture. There are $L$ terrestrial ISAC transmit APs (TX-APs) $\mathcal{L}=\{1, \cdots, L\}$ and $R$ terrestrial sensing receive APs (RX-APs) $\mathcal{R}=\{1, \cdots, R\}$, $K$ terrestrial single-antenna user equipments (UEs), and a surveillance area to detect unauthorized drones. All APs are equipped with $M$ antennas arranged in a horizontal uniform linear array (ULA) and interconnected through fronthaul links to a central cloud and fully synchronized \cite{cell-free-book}.  

A multi-location beam-searching method is employed, in which the surveillance area is divided into a set of non-overlapping sensing hotspots, $\mathcal{S} = \{\mathbf{q}_1, \ldots, \mathbf{q}_S\}$, where $|\mathcal{S}| = S$. $\mathbf{q}_i \in \mathbb{R}^3$ is the three-dimensional Cartesian coordinates of the sensing hotspot $i$. For each observation period, we select a subset of $N_S$ sensing hotspots. Thus, the entire area is scanned over $N_{\rm a}=\lceil\frac{S}{N_S}\rceil$ observation periods. We assume that for a given observation period, only one of the sensing hotspots might have an unauthorized drone.  

During a single observation period, the TX-APs cooperatively transmit orthogonal sensing beams toward multiple target locations. Simultaneously, a subset of RX-APs jointly sense these locations, assuming a line-of-sight (LOS) connection between each AP and its respective target. To ensure signal separation, we utilize a set of orthogonal sensing pilot sequences, allowing the received signals from different locations to remain orthogonal at the serving RX-APs. However, since the number of available pilots is limited to $T$, a single sensing pilot may be reused across multiple sensing locations when $T < N_S$.

Each TX-AP serves a set of UEs, denoted by $\mathcal{U}_l$. Simultaneously, the TX-APs contribute to sensing the surveillance area by transmitting an additional sequence of $\tau_s$ sensing symbols towards selected sensing hotspots. The sensing signal utilizes the same time-frequency resources as the communication signal, with $\tau_s$ denoting the sensing blocklength. 
Let $s_k[m]$ denote the zero-mean downlink communication symbol for UE $k$ at time instance $m$ with unit power, i.e., $\mathbb{E}\{|s_k[m]|^2\}=1$. At observation period $n$, we aim to sense simultaneously a set of sensing hotspots, $\mathcal{S}_n \subset \mathcal{S}$, where $\bigcup_n \mathcal{S}_n = \mathcal{S}$, and  $\mathcal{S}_n \bigcap \mathcal{S}_{n'} = \varnothing$ for all $n \neq n'$. Let $t_s \in \{1, \dots, T\}$ denote the index of the pilot assigned to the hotspot $s \in \mathcal{S}_n$. Note that since $T\leq |\mathcal{S}_n|=N_S$, several hotspots may share the same pilot signal. We let  $r_{t_s}[m]$ be the sensing symbol from pilot $t_s$ which is assigned to sensing hotspot $s$ at  time instance $m$ of observation period $n$. 
The transmitted signal $\textbf{x}_l[m] \in \mathbb{C}^M$ from TX-AP $l$ at time instance $m$ can be written as
\begin{equation}\label{x_k}
    \textbf{x}_l[m]= \sum_{k\in \mathcal{U}_l} \sqrt{\rho_{k,l}}\textbf{w}_{k,l} s_{k}[m]+\sum_{s\in \mathcal{T}_l}\sqrt{q_{s,l}}\boldsymbol{\omega}_{s,l} r_{t_s}[m], 
\end{equation}
where $\rho_{k,l}\geq 0$ and $q_{t,l}\geq 0$ are power control coefficients for UE $k$ and target $s$, respectively. $\mathcal{U}_l$ is the set of assigned UEs to TX-AP $l$ and $\mathcal{T}_l$ is the set of assigned sensing hotspots to TX-AP $l$ at the observation period $n$. 
At each TX-AP, we allocate a total power to sensing, $P_s$, and a total power to communication $P_c$ where $P_c+P_s=P_{\rm max}$. The sensing power is then divided equally between the assigned hotspots to that TX-AP. The communication power is distributed among the assigned UEs based on their channel gains, following the distributed power allocation algorithm in \cite[Section 7.2]{cell-free-book}. The power coefficients are then obtained as
\begin{align}
    \rho_{k,l}=\left\{\begin{matrix}
\frac{P_c\left(\Vert \textbf{h}_{k,l}\Vert^2\right)^v}{\sum_{l'\in \mathcal{U}_l}\left(\Vert \textbf{h}_{k,l}\Vert^2\right)^v}, & \quad k\in \mathcal{U}_l, \\
 0 & \quad  k\notin \mathcal{U}_l. \\
\end{matrix}\right.  
\end{align}
where $v=0.5$.
$\textbf{w}_{k,l}\in \mathbb{C}^{M}$ and $\boldsymbol{\omega}_{s,l}\in \mathbb{C}^M$ are the unit-norm transmit precoding vectors at TX-AP $l$ for UE $k$ and the sensing signal $t$, respectively, where $\sum_{k\in \mathcal{U}_l} \rho_{k,l} + \sum_{t\in \mathcal{T}_l}  q_{t,l} \leq P_{\rm{max}}$.

We use maximum ratio transmission (MRT) precoding vectors for both communication and sensing, given by $\textbf{w}_{k,l}=\frac{\textbf{h}_{k,l}}{\Vert \textbf{h}_{k,l}\Vert}$ and $\boldsymbol{\omega}_{s,l} = \frac{\textbf{a}(\varphi_{s,l},\vartheta_{s,l})}{\Vert \textbf{a}(\varphi_{s,l},\vartheta_{s,l})\Vert}$, respectively.
$\textbf{h}_{k,l}^*\in \mathbb{C}^M$ is the channel from TX-AP $l$ to UE $k$ and $\textbf{a}(\varphi_{s,l},\vartheta_{s,l})\in \mathbb{C}^M$ is the antenna array response vector corresponding to TX-AP $l$ with respect to target $s$ where $\varphi_{s,l}$ and $\vartheta_{s,l}$ are the azimuth and elevation angles from TX-AP $l$ to the target location $s$, respectively.
Assuming the antennas at the APs are half-wavelength-spaced,  
$\textbf{a}(\varphi_{s,l},\vartheta_{s,l}) =\begin{bmatrix}
          1& e^{j\pi \sin(\varphi_{s,l})\cos(\vartheta_{s,l})}& \ldots& e^{j(M-1)\pi\sin(\varphi_{s,l})\cos(\vartheta_{s,l})}
        \end{bmatrix} ^T$ \cite{bjornson2017massive}. 
\enlargethispage{-0.05in}

The received signal at UE $k$ is given as
\vspace{-1mm}
\begin{align}    \label{y_i}
     y_k[m] =&\underbrace{\sum_{l=1}^{L}\sqrt{\rho_{k,l}}\textbf{h}_{k,l}^{H}\textbf{w}_{k,l} s_{k}[m]}_{\textrm{Desired signal}}\nonumber\\
     &+ \underbrace{\sum_{l=1}^{L}\sum_{j=1,j\neq k}^{K}\sqrt{\rho_{j,l}}\textbf{h}_{k,l}^{H}\textbf{w}_{j,l} s_{j}[m]}_{\textrm{Interference signal due to  other UEs}}\nonumber\\
    &+ \underbrace{\sum_{l=1}^{L}\sum_{s\in \mathcal{S}_m}\sqrt{q_{t,l}}\textbf{h}_{k,l}^{H}\boldsymbol{\omega}_{s,l}r_{t_s}[m]}_{\textrm{Interference signal due to  sensing}}+ \underbrace{n_k[m]}_{\textrm{Noise}} ,
\end{align}
where $n_k[m] \sim \mathcal{CN}(0,\sigma_n^2)$ denotes the additive White Gaussian noise (AWGN) at UE $k$. The spectral efficiency (SE) of UE $k$ is given by
\begin{align}
    \mathsf{SE}_k= \log_2 (1+ \gamma_k)
\end{align}
where $\gamma_k$ is the effective SINR at UE $k$, and given by \eqref{sinr_i} on the top of next page. 
\begin{figure*} \centering
\begin{align}\label{sinr_i}
     \gamma_k 
     &= \frac{\sum_{l=1}^{L}\rho_{k,l}\left\vert \textbf{h}_{k,l}^H \textbf{w}_{k,l}\right\vert^2}{\sum_{l=1}^{L}\sum_{j=1,j\neq k}^{K}\rho_{j,l}\left\vert \textbf{h}_{k,l}^H \textbf{w}_{j,l}\right\vert^2+\sum_{l=1}^{L}\sum_{t=1}^{T}q_{t,l}\left\vert \textbf{h}_{k,l}^H \boldsymbol{\omega}_{t,l}\right\vert^2+\sigma_n^2}.
\end{align}
\end{figure*}
\section{Network Configuration for Aerial Surveillance} 
In this section, we present the proposed approaches for hotspot grouping, TX-APs, RX-APs, and pilot assignment for the sensing hotspots. 
\vspace{-1mm}
\subsection{Hotspot Grouping}
We group the sensing hotspots by selecting $N_s$ hotspots in each observation period, aiming to minimize the interference among them. To do that, we maximize the pairwise distances between hotspots within each group. We first rank distances from each sensing hotspot to other hotspots. Then, we find the subset $\mathcal{B}^*$ starting from the hotspot having the lowest sum-distance from $N_{\rm a}-1$ closest neighbors. Finally, we assign the hotspots to different groups, trying to avoid assigning neighbors to the same group. The steps of the algorithm are listed in Algorithm.~\ref{alg:grouping}. 

For each sensing hotspot, we select $T_{\rm s}$ TX-APs and $R_{\rm s}$ RX-APs with the highest channel gain.
For communication UEs, we first sort the TX-APs by their channel gains, then sequentially select APs until the sum channel gain exceeds a predefined threshold.

\begin{algorithm}[h]
\caption{Greedy Distance-Based Hotspot Grouping}
\label{alg:grouping}
\begin{algorithmic}[1]
\State \textbf{Input} Number of sensing hotspots $S$ and their locations $\mathcal{S} = \{\mathbf{q}_1, \ldots, \mathbf{q}_S\}$, and number of hotspot groups $N_{\rm a}$
\State Initialize empty groups $\{\mathcal{S}_1,\ldots,\mathcal{S}_{N_{\rm a}}\}$ and remaining set $\mathsf{R}=\{1,\ldots,S\}$.
\State Compute Euclidean distance matrix $D$:
\Statex \quad $D(s,s') \gets \Vert \textbf{q}_s -\textbf{q}_{s'}\Vert, \quad \forall s,s'$
\While{$|\mathsf{R}| > 0$}
    \If{$|\mathsf{R}| \le N_{\rm a}$}
        \State Assign each remaining point to a distinct group;
        \State \textbf{break}
    \EndIf
\State For each hotspot rank distances to other hotspot.
\State Find the subset $\mathcal{B}^*$ starting from hotspot having the lowest sum-distance from $N_{\rm a}-1$ closest neighbors.
\Statex \quad $\mathcal{B}^* = \arg\min_{\mathcal{B}\subseteq\mathsf{R}, |\mathcal{B}|=G} \sum_{s,s'\in\mathcal{B}} D(s,s's)$
\State Distribute each one to another time group.

\State Assign subset points:
    \For{$x \in \mathcal{B}^*$}
        \State Define neighborhood $\mathcal{N}(x)=\{x\pm1, x\pm9, x\pm10, x\pm11\}\cup(\mathcal{B}^*\setminus\{x\})$
        \For{$n=1:N_{\rm a}$}
            \If{$\mathcal{S}_n \cap \mathcal{N}(x) = \emptyset$ \textbf{or} $n=N_{\rm a}$}
                \State $\mathcal{S}_n \gets \mathcal{S}_n \cup \{x\}$; \textbf{break}
            \EndIf
        \EndFor
    \EndFor
    \State $\mathsf{R} \gets \mathsf{R} \setminus \mathcal{B}^*$
\EndWhile
\State \textbf{Output:} Groups $\{\mathcal{S}_1,\ldots,\mathcal{S}_{N_{\rm a}}\}$
\end{algorithmic} \vspace{-1mm}
\end{algorithm}


\vspace{-1mm}
\subsection{Sensing Pilot Assignment}
In the sensing pilot assignment, an ambiguity parameter is defined to assess how much the target to RX-AP channels will resemble for two different hotspots. The concatenated channel regarding the sensing hotspot $s$ becomes 
\begin{align}  
\bar{\textbf{h}}_s = [\sqrt{\beta_{1,s}}\textbf{a}^T(\varphi_{s,1},\vartheta_{s,1}), \ldots, \sqrt{\beta_{R,s}}\textbf{a}^T(\varphi_{s,R},\vartheta_{s,R})]^T.
\end{align}
Let $\boldsymbol{\varrho}_s = [\varrho_{s,1}, \ldots, \varrho_{s,R}]^T$ denote the binary RX-AP clustering vector, and  $\varrho_{s,r}=1$ if $r$th RX-AP is assigned for the $s$th hotspot, and $\varrho_{s,r}=0$ otherwise. In the hypothesis test, only the samples from the chosen RX-APs for the $s$th hotspot will be used. Therefore, to assess an ambiguity between two hotspot channels, we should consider only the union of the set of receivers for these hotspots. For the hotspot $s$, the updated channel vector considering only the RX-APs of $s$ and $s'$ becomes
\begin{equation}
    \tilde{\textbf{h}}_{s|s'} = \operatorname{max}\{\boldsymbol{\varrho}_s, \boldsymbol{\varrho}_{s'}\}  \bar{\textbf{h}}_{s}.
\end{equation}

Then the ambiguity parameter between the sensing hotspots $s$ and $s'$ can be written as 
\begin{equation}
    \zeta_{ss'}  = \frac{|\tilde{\textbf{h}}_{s'|s}^H \tilde{\textbf{h}}_{s|s'}|}{\|\tilde{\textbf{h}}_{s|s'}\|^2_2},  
\end{equation}
where $\zeta_{ss'} \in [0,1]$. We aim to assign the same pilots to a group of sensing hotspots with a lower ambiguity parameter. However, we will have several pilot groups, where minimizing the sum ambiguity in one group might result into maximizing another group. To ensure balanced performance between hotspots, we minimized the maximum sum ambiguity of a pilot group. We let $\alpha_{st} \in \{0,1\}$ denote the membership of hotspot $s$ in pilot $t$, and $\boldsymbol{\alpha}_t = [\alpha_{1t}, \ldots, \alpha_{N_St}]^T$. We also concatenate all ambiguity parameter into a matrix, $\boldsymbol{\zeta} \in \mathbb{R}^{N_S\times N_S}$, where $[\boldsymbol{\zeta}]_{s,s'} = \zeta_{ss'} $ and diagonal entries being zero. 

We define a  binary matrix, $\textbf{A}_t = \boldsymbol{\alpha}_t\boldsymbol{\alpha}^{T}_t \in \left\{0, 1\right\}^{N_S \times N_S}$ and $[\textbf{A}_t]_{s,s'}= a_{s,s',t}= \alpha_{s,t} \alpha_{s',t}$.  The following constraints can replace the elements of the matrix:
\begin{equation}
    \begin{aligned}
&a_{s,s',t} \leq \alpha_{s,t}, \quad a_{s,s',t} \leq \alpha_{s',t}, \quad
 a_{s,s',t} \geq \alpha_{s,t}+\alpha_{s',t}-1,
\label{eq:constraint:binary_matrix}
\end{aligned}
\end{equation} 
where $a_{s,s',t}  \in \{0,1\}$. The optimization problem is given as
\begin{subequations} \label{eq:group_opt1:problem}
\begin{align}
 & \underset{\{\alpha_{s,t}, a_{s,s',t} \},\varsigma}{\text{minimize}}  \quad \varsigma \label{eq:group_opt1:objective} \\ & \textrm{subject to} \quad \eqref{eq:constraint:binary_matrix} \quad \nonumber \\ &
 \operatorname{Tr}( \boldsymbol{\zeta}{A}_t)
 \leq \varsigma, \quad \forall t
 \\ &
 \sum_{t=1}^{T} \alpha_{s,t} = 1, \quad \forall s, \quad \sum_{s=1}^{N_S} \alpha_{s,t} \leq \lceil N_S/T \rceil, \quad \forall t  \label{eq:group_opt:limitgroupsize} \\ &
 \alpha_{s,t}, a_{s,s',t} \in \{0,1\}, \quad \forall s,s',t. \label{eq:group_opt1:binary}
 \end{align}
\end{subequations}
where \eqref{eq:group_opt1:objective} 
aims to minimize the maximum sum ambiguity in a group of hotspots sharing the same pilot, denoted by $\varsigma$. \eqref{eq:group_opt:limitgroupsize} ensures that a hotspot is only assigned to a single pilot, and the total number of hotspots sharing a pilot is limited. 
Global optimum can be obtained for this problem by using a branch-and-bound algorithm, which in this work is implemented by MOSEK with CVX in MATLAB.

\section{Target Detection}
We consider multi-static sensing, meaning that the sensing transmitters and the receivers are not co-located. We assume that the target-free channel between TX-AP $l$ and RX-AP $r$ is acquired prior to sensing in the absence of the target. The transmit signal $\textbf{x}_l[m]$ is also known at the central cloud. Hence, except the noise, the undesired part of the received signal at each RX-AP can be canceled. We employ a distributed maximum ratio combining (MRC) scheme at each RX-AP $r\in\mathcal{R}$, where the received signal is combined using $\boldsymbol{v}_{r,s}^H=\frac{\textbf{a}^H(\phi_{r,s},\theta_{r,s})}{\|\textbf{a}(\phi_{r,s},\theta_{r,s})\|}$. In addition, matched filtering is applied by correlating the combined signal with the Hermitian of pilot sequence $\mathbf{r}_{t_s}^H$. In the presence of the target, RX-AP $r$ assigned to hotspot $s$ receives either the desired target's reflections from hotspot $s$ or the interference due to target's reflections from hotspot $s'\neq s$, depending on the location of the target. Let us define $\mathbb{I}(s)\in \{0,1\}$ representing the presence/absence of the target in hotspot $s$, where $\mathbb{I}(s)\neq \mathbb{I}(s')$ and $\sum_{s=1}^{S}\mathbb{I}(s)=1$. Thus, the received signal at AP $r\in\mathcal{R}$ over $\tau_s$ symbols corresponding to the sensing hotspot $s$ with pilot sequence $\textbf{r}_{t_s}$ is \footnote{In practice, one should also take the cancellation error into account, which is left as future work.}
\enlargethispage{-0.1in}
\vspace{-2mm}
\begin{align}\label{y_r}
         & z_{r,s} 
          = \sqrt{M} \tau_s \boldsymbol{\beta}_{r,s}^T\boldsymbol{\alpha}_{r,s}\mathbb{I}(s)\nonumber\\
          & + \tau_s \sum_{s'\in \mathcal{S}_{t_s}\setminus s} \frac{\textbf{a}^H(\phi_{r,s},\theta_{r,s})\textbf{a}(\phi_{r,s'},\theta_{r,s'})}{\|\textbf{a}(\phi_{r,s},\theta_{r,s})\|}\boldsymbol{\beta}_{r,s'}^T\boldsymbol{\alpha}_{r,s'}\mathbb{I}(s')+n'_{r,s}    
\end{align}
where $n'_{r,s}\sim \mathcal{CN}(0,\sigma_n^2\tau_s)$ is the combined receive noise.
For notational simplicity, we have defined
$\boldsymbol{\beta}_{r,s} \in \mathbb{C}^{L}$ as 
\begin{align}
  \boldsymbol{\beta}_{r,s}=  \begin{bmatrix}\sqrt{\beta_{r,s,1}} \textbf{a}^{T}(\varphi_{1,s},\vartheta_{1,s})\sqrt{q_{s,1}}\boldsymbol{\omega}_{s,1}\\\dots\\ \sqrt{\beta_{r,s,L}}  \textbf{a}^{T}(\varphi_{L,s},\vartheta_{L,s})\sqrt{q_{s,L}}\boldsymbol{\omega}_{s,L}\end{bmatrix}
\end{align} 
and $\boldsymbol{\alpha}_{r,s} \triangleq  \begin{bmatrix}
\alpha_{r,s,1}&\alpha_{r,s,2}&\ldots&\alpha_{r,s,L}\end{bmatrix}^T\in \mathbb{C}^{L}$. 
We form the concatenated received signal $\textbf{z}_s[m]\in \mathbb{C}^{R}$ by all $R$ RX-APs involved in the sensing, i.e., $r\in\mathcal{R}$, as follows
\begin{align} 
\label{z_rPrim}
          \textbf{z}_s
          &=  \sqrt{M} \tau_s\boldsymbol{\beta}_s^T \boldsymbol{\alpha}_s \mathbb{I}(s)+ \tau_s\!\!\! \sum_{s'\in \mathcal{S}_{t_s}\setminus s} \!\!\! \boldsymbol{\mathcal{A}}_{ss'}\boldsymbol{\beta}_{s'}^T\boldsymbol{\alpha}_{s'}\mathbb{I}(s')+ \!\textbf{n}_s,
\end{align}
where $\boldsymbol{\mathcal{A}}_{ss'}\in\mathbb{C}^{R\times R}$ is a diagonal matrix with $[\boldsymbol{\mathcal{A}}_{ss'}]_{r,r}= \frac{\textbf{a}^H(\phi_{r,s},\theta_{r,s})\textbf{a}(\phi_{r,s'},\theta_{r,s'})}{\|\textbf{a}(\phi_{r,s},\theta_{r,s})\|}$, $\boldsymbol{\beta}_s^T= \operatorname{bdiag}\left(\boldsymbol{\beta}_{1,s}^T,\ldots, \boldsymbol{\beta}_{R,s}^T\right)\in \mathbb{C}^{R\times R\,L}$, $\boldsymbol{\alpha}= \begin{bmatrix}\boldsymbol{\alpha}_1^T & \ldots & \boldsymbol{\alpha}_{R}^T
\end{bmatrix}^T\in\mathbb{C}^{R\,L}$, and $\textbf{n}[m] = \begin{bmatrix}
    n_1'[m]& \ldots& n_{R}'[m]
\end{bmatrix}^T \in \mathbb{C}^{R}$.

Given that the ambiguity among the hotspots is negligible after the hotspot groping and pilot assignment, the hypotheses to detect the target can be written as follows:
\begin{align}
    &\mathcal{H}_0 : \textbf{z}_s= \textbf{n}_s,
    \\
    &\mathcal{H}_1 : \textbf{z}_s= \sqrt{M} \tau_s\boldsymbol{\beta}_s^T \boldsymbol{\alpha}_s+\textbf{n}_s[m],
\end{align}
where  $\mathcal{H}_0 $ represents the hypothesis that there is no target and   $\mathcal{H}_1 $ represents the hypothesis that the target exists and the reflected signals from the target are received by the RX-APs. Note that, in our numerical analysis, we consider the ambiguity between hotspots, even though the detector itself remains unaware of it.  Employing the maximum a posteriori ratio test (MAPRT) detector, the test statistic is \cite{wcnc2025UAV} 
\begin{align}
    T = \textbf{z}_s^H \mathbf{B}_s \textbf{z}_s\label{eq:MAPRT}
\end{align}
where $
    \mathbf{B}_s = M^2 \boldsymbol{\beta}_s^T \left(M^2\tau_s \boldsymbol{\beta}_s^* \boldsymbol{\beta}_s^T + \sigma_n^2 \textbf{I}_{RL}\right)^{-1} \boldsymbol{\beta}_s^*$. 
Finally, the true hypothesis $\hat{\mathcal{H}}$ is estimated as 
\begin{align}
    \hat{\mathcal{H}} = \left\{\begin{matrix}
\mathcal{H}_0, & \textrm{if}\quad T<\lambda, \\
\mathcal{H}_1, & \textrm{if} \quad T\geq\lambda, \\
\end{matrix}\right.
\end{align}
where $\lambda$ is the detector threshold and its value is set to meet a given false alarm probability threshold. 

\section{Age of Sensing (AoS) }
AoS can be defined as the elapsed time from the last sensing update. We proposed a deterministic sensing approach, where the network periodically updates the sensing information. In this approach, the target detection decision for a given hotspot is updated once after sensing the whole surveillance area over all time groups. Thus, AoS for our system can be modeled deterministically as
\begin{align}
\Delta_{\mathrm{total}} = \sum_{n=1}^{N_a} T_{n}=\sum_{n=1}^{N_a} (T_{n}^{\rm tr}+T_{n}^{\rm pr}+T_{n}^{\rm net}),
\end{align}
where $T_n$ is the total time required to sense all sensing hotspots in the $n^{\rm th}$ observation period, including transmission delay $T_n^{\rm th}$, processing delay $T_n^{\rm pr}$, and network-related delay $T_n^{\rm net}$. The transmission delay takes into account of the time required to observe locations-- including transmitting and receiving sensing signals at the APs-- and forward the combined signals to the cloud, given as
\begin{align}
    T_{n}^{\rm tr} &= D_n^{\rm obs}+D_{n}^{\rm F,tx}+ D_{n}^{\rm F,rx}  +D_{s}^{\rm R,tx}+D_{s}^{\rm R,rx}
\end{align}
where $D_n^{\rm obs}= \frac{\tau_s }{B}+ D^{\rm p}_{\rm 2-way}$ is the observation delay including the transmission and reception delay at the TX-AP which is a function of sensing blocklength and bandwidth (i.e., $ \frac{\tau_s }{B}$) and two-way propagation delay corresponding to TX-target-RX path. $D_{s}^{\rm F,tx}$ and $ D_{s}^{\rm F,rx}$ are the transmission and reception delay in the fronthaul link given by \cite{8479363}
\begin{align}
    D_{s}^{\rm F,x}= \lceil \frac{\textrm{FSL}_{\rm x}N_{\rm bit}}{L_{\rm packet}}\rceil. \frac{L_{\rm packet}}{R_{\rm L}}, \quad \textrm{x}= \textrm{tx, rx}
\end{align}
where $\textrm{FSL}_{\rm x}$ is the number of real scalars should be transmitted from the cloud to the TX-AP or the RX-AP to the cloud, listed in Table~\ref{tab:FSL} where $\vert \mathcal{T}_{l}\vert$ and $\vert \mathcal{R}_{r}\vert$ are the number of sensing hotspots assigned to TX-AP $l$ and RX-AP $r$, respectively.  $L_{\rm packet}$ is the Ethernet frame size in bits and $N_{\rm bit}$ is the number of bits per scalar (I or Q) sample which is equal to $15$ for CPRI (Common Public Radio Interface)\cite{holma20245g}.
\begin{table}

 \caption{Fronthaul Signaling Load Per AP ($\textrm{FSL}_{\rm x}$)}\label{tab:FSL}
    \centering
    \begin{tabular}{|c|c|c|}
    \hline
    \multirow{1.5}{*}{ Sensing, TX}&\multirow{1.5}{*}{ Communication, TX}&\multirow{1.5}{*}{ Sensing, RX}\\[2mm]\hline
    \multirow{1.5}{*}{ $2\tau_{\rm s} \vert \mathcal{T}_{l}\vert$}&\multirow{1.5}{*}{ $2\tau_{\rm s} \vert \mathcal{U}_{l}\vert$}&\multirow{1.5}{*}{ $\vert \mathcal{R}_{r}\vert$ }\\[2mm]
        \hline     
    \end{tabular}
   \vspace{-3mm}
\end{table}
$D_{s}^{\rm R,tx}$ and $D_{s}^{\rm R,rx}$ are the transmission and reception delay in the RAN, respectively and they are fixed depending on the distance. 

The processing delay is given as 
\begin{align}
    T_{n}^{\rm pr} &=D_{n}^{\rm pr,C} +D_{n}^{\rm pr,AP} \nonumber\\
    &= \frac{C_{\rm sens}^{\rm C}}{C_{\rm max,sense}^{\rm C}}+\frac{C_{\rm sense}^{\rm txAP}}{C_{\rm max,sense}^{\rm AP}}+\frac{C_{\rm sense}^{\rm rxAP}}{C_{\rm max,sense}^{\rm AP}} 
\end{align}
where $D_{s}^{\rm pr,c}$ and $D_{s}^{\rm pr,AP}$ are the processing delay at the cloud and the APs, respectively. $C_{\rm sens}^{\rm C}$ and $C_{\rm sens}^{\rm AP}$ are the number of giga-operations (GOP) corresponding to  sensing processing at the cloud and APs, respectively. The maximum allocated resources to sensing at the cloud is $C_{\rm max,sense}^{\rm C}$ and the AP is $C_{\rm max,sense}^{\rm AP}$ in giga-operation per second (GOPS). We calculate the number of real multiplications and divisions. Each complex multiplication costs 4 real multiplications. We also consider memory overhead in arithmetic operation calculations by multiplying each operation by two as done in \cite{demir2023cell}. Hence, each complex multiplication is counted as $4\cdot 2= 8$ operations.  
The GOP calculation at the TX-APs and RX-APs are 
\begin{align}
    C_{\rm sense}^{\rm txAP}&=\frac{1}{ 10^{9}}12 M \tau_s\vert \mathcal{T}_{l}\vert,\\
    C_{\rm sense}^{\rm rxAP}&= \frac{1}{ 10^{9}} \left(8M\tau_s\vert \mathcal{R}_{r}\vert + 8\tau_s\vert \mathcal{R}_{r}\vert\right),
\end{align}
where the details are listed in Table~\ref{tab:gops_AP}. Computing each test statistics at the cloud costs $\left( 8 R_{\rm s}^2+8  R_{\rm s}\right)$ operations. Thus, the total GOP for each observation period is
\begin{align}
    C_{\rm sens}^{\rm C}= \frac{\vert \mathcal{S}_{\rm m}\vert}{ 10^{9}} \left( 8 R_{\rm s}^2+8  R_{\rm s}\right).
\end{align}
Note that we neglect the computation complexity of computing the matrices $\textbf{B}_s$ since we only need to compute them once and they will be fixed over all observation periods. 

\begin{table}
 \caption{Sensing GOP per AP}\label{tab:gops_AP}
    \centering
    \begin{tabular}{|c|c|}
    \hline
    \multirow{1.5}{*}{ Operation} &\multirow{1.5}{*}{ Computational Complexity} \\[2mm]
    \hline
        \multirow{1.5}{*}{ MR Precoding multiplication} &\multirow{1.5}{*}{$8\tau_sM\vert \mathcal{T}_{l}\vert$  } \\[2mm]
        \hline
         \multirow{1.5}{*}{DL Power multiplication}& \multirow{1.5}{*}{ $4 \tau_s M \vert \mathcal{T}_{l}\vert$} \\[2mm]
         \hline
        \multirow{1.5}{*}{ MR combining vector multiplication }& \multirow{1.5}{*}{$8\tau_s M\vert \mathcal{R}_{r}\vert$ }\\[2mm]
         \hline
         \multirow{1.5}{*}{ Pilot multiplication ($\textbf{r}_{t_s}^H$) }& \multirow{1.5}{*}{$8\tau_s\vert \mathcal{R}_{r}\vert$ }\\[2mm]
         \hline
    \end{tabular}
\vspace{-3mm}
\end{table}

\section{Numerical Results}
\begin{figure}[t!]
    \centering
   \includegraphics[width=0.9\linewidth]{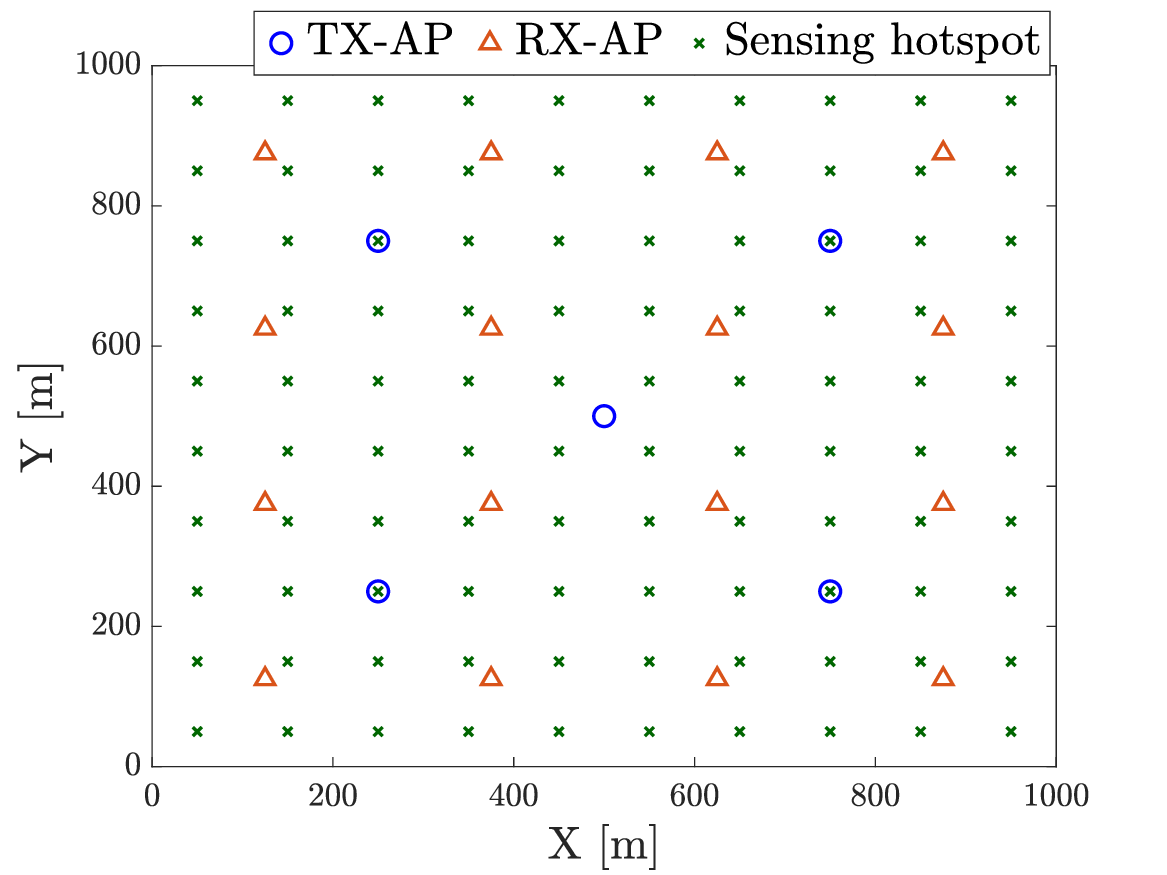} 
   \caption{A helicopter view of the simulation scenario. }
   \label{fig:locations}
    \vspace{-4mm}
   \end{figure}
We consider an area of $1000\times 1000\,m^2$ with $5$ ISAC TX-APs, $16$ sensing RX-APs, $S=100$ sensing hotspots, as shown in Fig.~\ref{fig:locations}. There are $K=8$ single-antenna UEs randomly located in the area. Each AP is equipped with $M=16$ antenna elements. 
The number of orthogonal pilots is set to $5$. Maximum transmit power per AP is $P_{\rm max}=1$\,Watt and the RCS of the target is $-30$\,dBsm. False alarm probability and detection probability thresholds are $0.03$ and $0.9$, respectively.  
$D^{\rm p}_{\rm 2-way}$ is the maximum two-way propagation delay in the system. Maximum allocated processing resource at the cloud and APs is $10$ GOPS. The rest of the parameters are listed in Table.~\ref{tab:simulation}.

\begin{table}
 \caption{Simulation Parameters}
    \centering\label{tab:simulation}
    \begin{tabular}{|c|c|c|c|c|}
    \hline
    \multirow{1.5}{*}{  Parameter}&\multirow{1.5}{*}{$D^{\rm R,tx},D^{\rm R,rx}$ } &\multirow{1.5}{*}{ $N_{\rm bit}$} &\multirow{1.5}{*}{$L_{\rm packet}$ } &\multirow{1.5}{*}{$T_n^{\rm net}$ }\\[2mm]
         \hline \multirow{1.5}{*}{ Value}& \multirow{1.5}{*}{$5\mu s$ }&\multirow{1.5}{*}{ $15$ bits}& \multirow{1.5}{*}{$1500$ Bytes }& \multirow{1.5}{*}{$60\,\mu\text{s}$}
         \\[2mm]
         \hline
    \end{tabular}
\vspace{-4mm}   
    \label{tab:placeholder}
\end{table}

\begin{figure}[t!]
    \centering
   \includegraphics[width=0.9\linewidth]{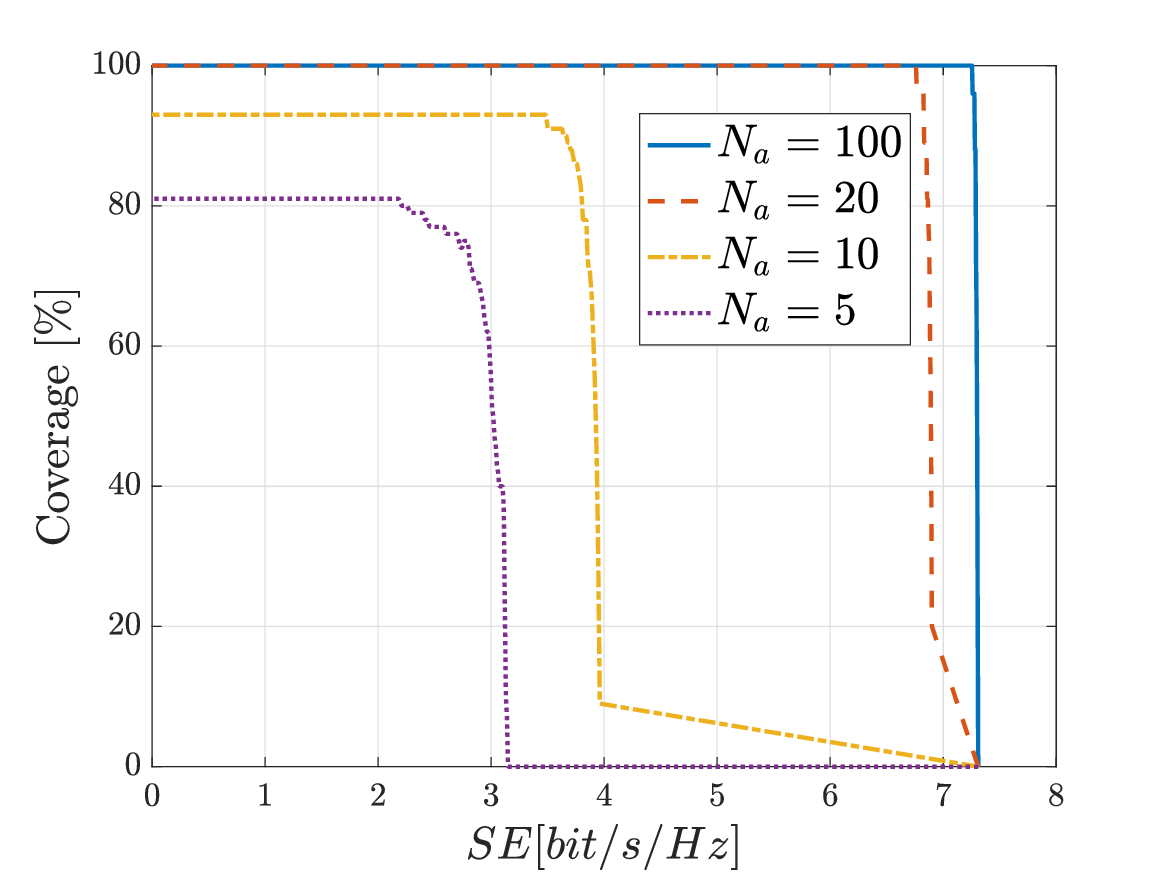} 
   \caption{Sensing coverage vs. communication SE per UE. }
   \label{fig:cov_SE}
   \vspace{-4.5mm}
   \end{figure}

\begin{figure}[t!]
    \centering
   \includegraphics[width=0.9\linewidth]{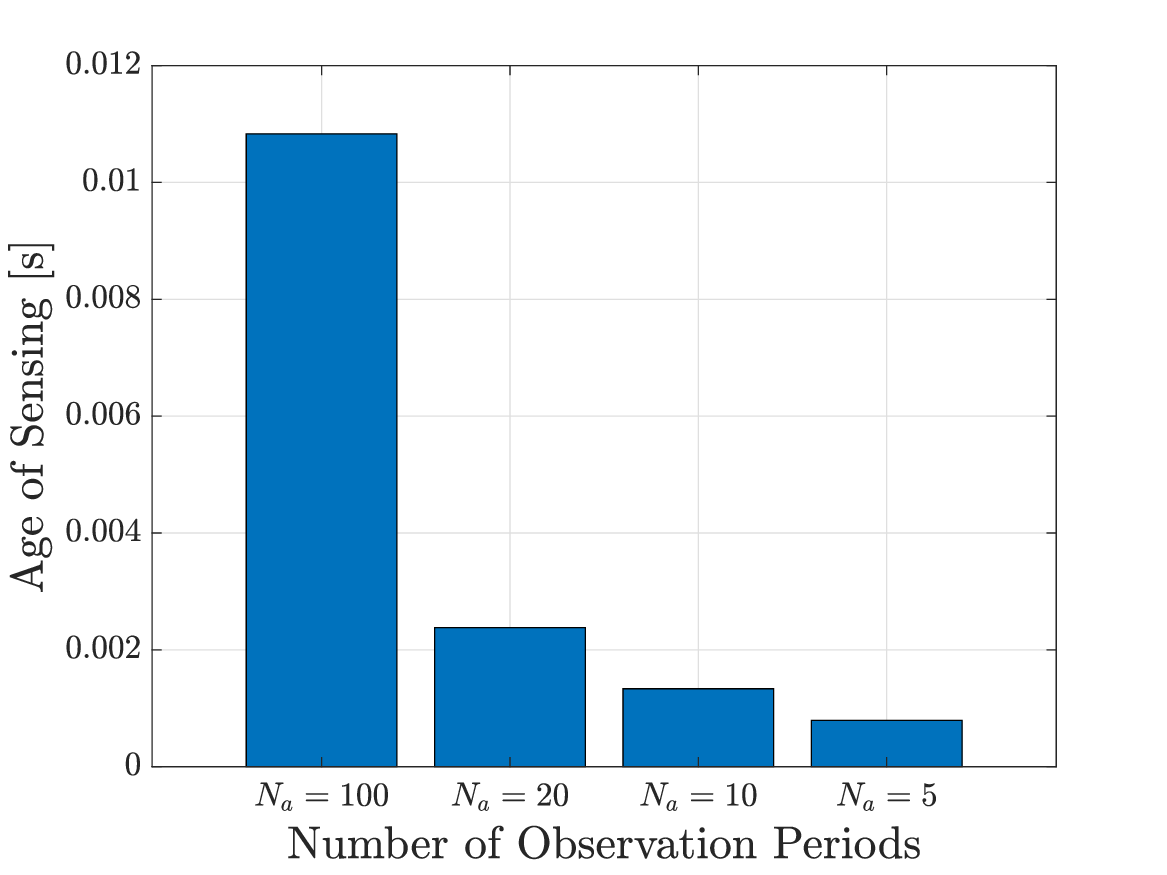} 
   \caption{Age of Sensing vs number of hotspot groups. }
   \label{fig:age_Na} \vspace{-4.5mm}
   \end{figure}

Fig.~\ref{fig:cov_SE} illustrates the trade-off between sensing coverage and SE per UE for different numbers of observation periods. The case $N_a=100$ corresponds to a configuration without hotspot grouping, where each sensing hotspot is observed individually within a single period. In contrast, $N_a=20, 10$ and $5$ represent the cases with hotspot grouping. In general, sensing coverage remains relatively stable at low SE values. However, beyond a certain point, coverage drops sharply as SE increases. This behavior occurs because achieving higher SE requires allocating more power to communication, thereby reducing the power available for sensing. Furthermore, as the number of observation periods decreases, more hotspots are grouped within each period, which not only reduces the per-hotspot sensing power but also increases hotspot ambiguity—particularly for $N_a=10$ and $5$. As a result, the detection probability decreases, leading to reduced overall sensing coverage. In addition, reducing $N_a$ also lowers the achievable communication SE due to the increased interference generated by the sensing signals. Nevertheless, the results indicate that full sensing coverage ($100\%$) can still be achieved with only a
$6.9\%$ reduction in SE compared to the $N_a=100$ case.
Fig.~\ref{fig:age_Na} shows the AoS for different numbers of observation periods $N_a=100, 20, 10$ and $5$. The results show that hotspot grouping can significantly reduce the AoS. For instance, when $N_a=20$, the AoS decreases by up to $78.7\%$ compared to when $N_a=100$ and up to $92.6\%$ with $N_a=5$.

\begin{figure}[t]
\begin{subfigure}{0.53\linewidth}
  \centering
  \includegraphics[trim={0mm 0mm 0mm 0mm},clip,width=\linewidth]{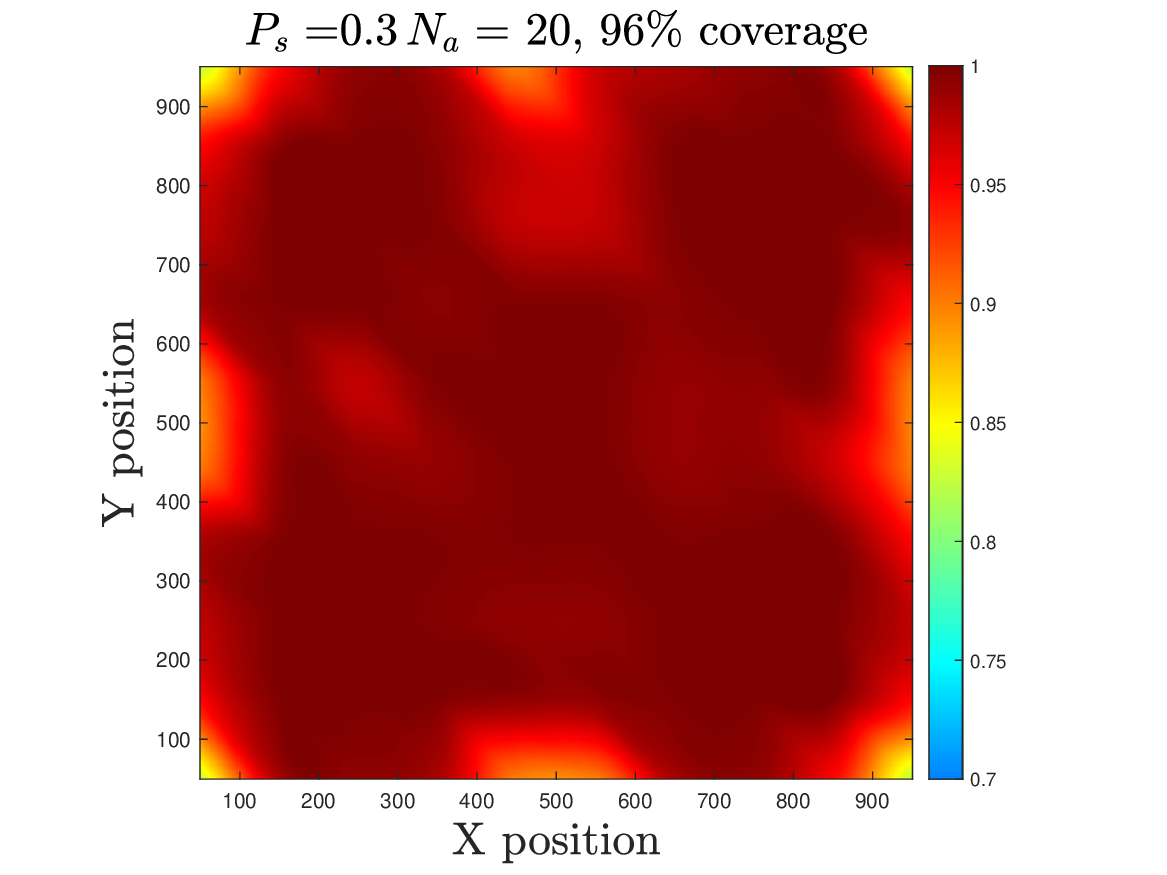} \caption{No ambiguity}
    \label{figure:no ambiguity}
    \end{subfigure}
  \hspace{-1cm}
\begin{subfigure}{0.53\linewidth}
  \centering
  \includegraphics[trim={0mm 0mm 0mm 0mm},clip,width=\linewidth]{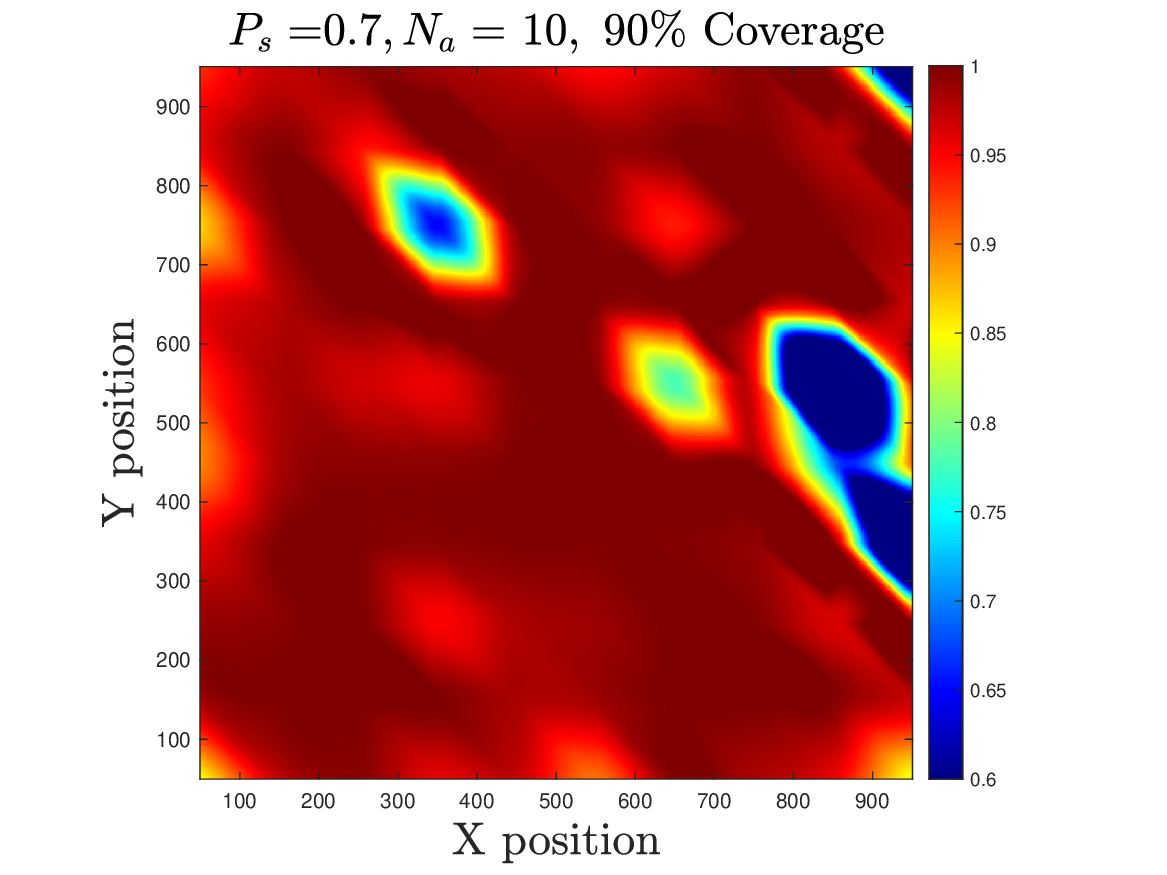} \caption{Negligible ambiguity}
   \label{figure:negligible ambiguity}
    \end{subfigure}
    \vfill
       \hspace{0.17\linewidth}
    \begin{subfigure}{0.53\linewidth}
  \centering
  \includegraphics[trim={0mm 0mm 0mm 0mm},clip,width=\linewidth]{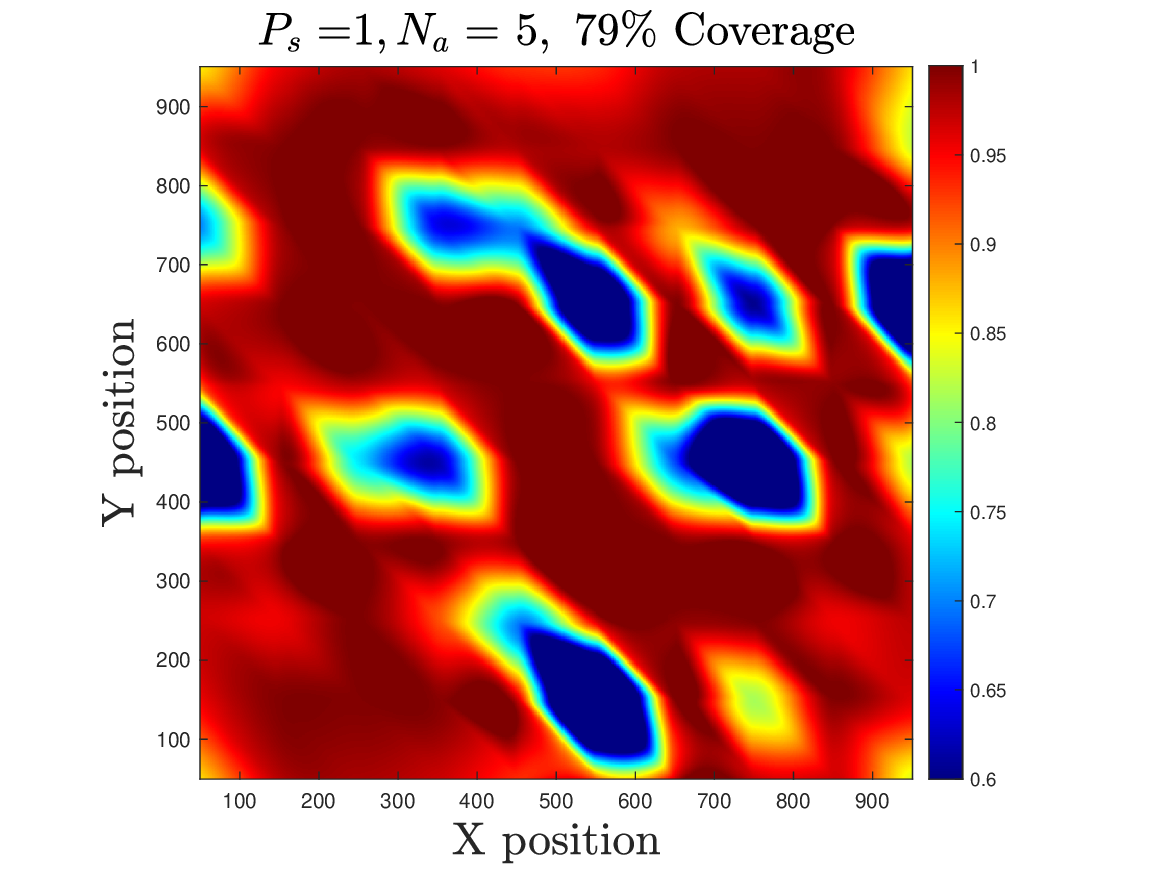} \caption{Low ambiguity}
   \label{figure:low ambiguity}
    \end{subfigure}
    \vspace{-2mm}
   \caption{Heatmap sensing coverage for a) $P_s=0.3$ and $N_a=20$, b)$P_s=0.7$ and $N_a=10$, and c)$P_s=1$ and $N_a=5$. }\label{fig:heatmap}
   \vspace{-2mm}
\end{figure}

Fig.~\ref{fig:heatmap}a–c illustrate the sensing coverage as a function of total sensing power per AP and the number of observation periods. The results indicate that coverage decreases as the number of observation periods is reduced, primarily due to increased ambiguity, even when the total sensing power per AP is increased. For instance, when $N_a=20$, each observation period covers $5$ sensing hotspots. With $5$ sensing pilots, no hotspots share a pilot, resulting in zero ambiguity. However, as the number of observation periods decreases, more hotspots share the same pilot, leading to higher ambiguity and lower coverage. Additionally, the total sensing power at each TX-AP is distributed among more hotspots, which further degrades performance.

\vspace{-2mm}
\section{Conclusion}
In this paper, we propose a novel cell-free ISAC framework for detecting unauthorized drones using multi-point sensing. We formulate a closed-form expression for the AoS to quantify the timeliness of sensing information. Furthermore, we introduce a hotspot ambiguity parameter and propose a comprehensive network configuration strategy, including hotspot grouping, AP clustering, and sensing pilot assignment, to reduce AoS while maintaining high coverage. Our results demonstrate that grouping hotspots to minimize the number of observation periods, while keeping hotspot ambiguity limited, achieves the best trade-off between AoS and sensing coverage.   

\vspace{-2mm}
\bibliographystyle{IEEEtran}
\bibliography{IEEEabrv,refs}
\end{document}